\documentclass[manuscript]{aastex}

\begin{document}


\title{Pre-discovery observations of CoRoT-1b and CoRoT-2b with the BEST survey}


\author{H. Rauer\altaffilmark{1}, A. Erikson, P. Kabath,  P. Hedelt}
\affil{DLR, Institut f\"ur Planetenforschung, Rutherfordstr. 2,
12489 Berlin, Germany }

\author{M. Boer}
\affil{Observatoire de Haute Provence, St Michel-l'Observatoire, 04870,
France}

\author{L. Carone}
\affil{Rheinisches Institut f\"ur Umweltforschung, Universit\"at zu K\"oln,
Abt. Planetenforschung, Aachener Str. 209, 50931 K\"oln, Germany}

\author{Sz. Csizmadia}
\affil{DLR, Institut f\"ur Planetenforschung, Rutherfordstr. 2,
12489 Berlin, Germany}

\author{P. Eigm\"uller}
\affil{Th\"uringer Landessternwarte Tautenburg, Sternwarte 5, 07778
Tautenburg, Germany}

\author{P. v. Paris, S. Renner\altaffilmark{2}}
\affil{DLR, Institut f\"ur Planetenforschung, Rutherfordstr. 2, 12489 Berlin, Germany}

\author{G. Tournois}
\affil{Observatoire de Haute Provence, St Michel-l'Observatoire, 04870, France}

\and

\author{R. Titz, H. Voss\altaffilmark{3}}
\affil{DLR, Institut f\"ur Planetenforschung, Rutherfordstr. 2, 12489 Berlin, Germany}

\altaffiltext{1}{TU Berlin, Zentrum f\"ur Astronomie und Astrophysik,
Hardenbergstr. 36, 10623 Berlin, Germany}
\altaffiltext{2}{Laboratoire d'Astronomie de Lille, Universit\'e de Lille 1, 1 impasse de l'Observatoire, 59000 Lille, France}
\altaffiltext{3}{Departament d'Astronomia i Meteorologia, Facultat de F\'isica, Universitat de Barcelona, C/ Mart\'i i Franqu\`es 1, 08028 Barcelona, Spain}


\begin{abstract}
The BEST wide-angle telescope installed at the Observatoire de
Haute-Provence and operated in remote control from Berlin by the Institut f\"ur
Planetenforschung, DLR, has observed the  CoRoT target fields prior to the
mission. The resulting archive of stellar photometric lightcurves is used to
search for deep transit events announced during CoRoT's alarm-mode to aid in
fast photometric confirmation of these events. The "initial run" field of
CoRoT (IRa01) has been observed with BEST in November and December 2006 for 12
nights. The first "long run" field (LRc01) was observed from June to September
2005 for 35 nights. After standard CCD data reduction, aperture photometry has
been performed using the ISIS image subtraction method. About 30,000 lightcurves
were obtained in each field. Transits of the first detected planets by the
CoRoT mission, CoRoT-1b and CoRoT-2b, were found in archived data of the BEST
survey and their lightcurves are presented here. Such detections provide useful
information at the early stage of the organization of follow-up observations of
satellite alarm-mode planet candidates. In addition, no period change was
found over $\sim$4 years between the first BEST observation and last available
transit observations.
\end{abstract}

\keywords{Techniques: photometry - Planets and satellites: CoRoT-1b, CoRoT-2b}


\section{Introduction}

The Berlin Exoplanet Search Telescope (BEST) is a 19.5 cm aperture
wide angle telescope dedicated to time-series photometric
observations \cite{Rauer2004}. The main
purpose of the instrument is to provide ground-based support to the
CoRoT space mission (CNES; \cite{Baglin2006}). The mission target
fields are observed at least one year prior to the observations of the spacecraft.
Therefore, planetary transit candidates found by CoRoT at bright stars  can be searched for quickly in
the BEST archive to confirm the transit event and to check the
ephemeris. In addition, BEST data sets are used to search for new
variable stars in the CoRoT target fields which are input to the
additional science program, like e.g. eclipsing binary stars
(\cite{Karoff2007}, \cite{Kabath07}, \cite{Kabath08}). Furthermore, information on variable
stars nearby CoRoT transit candidates can be used to help
disentangling crowding problems during the CoRoT lightcurve
analysis.

The transit signals of the first two planets detected by CoRoT
(CoRoT-1b and CoRoT-2b) (\cite{barge2008}, \cite{alonso2008}), were found using the "alarm detection mode" of the
mission, during ongoing observations of the respective
target fields. Since both planets orbit relatively bright stars, we searched for signatures
of the transit candidates in our BEST data archive. Partial transit events of the planets were recorded by BEST in December 2006 and summer 2005 during observations of the CoRoT "initial run" and first "long run" fields.
In the following, we describe these pre-discovery observations of CoRoT-1b and CoRoT-2b.

\section{The BEST System}

\subsection{Telescope and Instrumentation}

The observations were performed with the Berlin Exoplanet
Search Telescope (BEST) system \cite{Rauer2004}. The BEST Telescope is a commercial flatfield telescope with
$19.5$ cm aperture and $f/2.7$ focal ratio and is mounted on a
German equatorial mount. The system uses an air cooled Apogee-10 CCD
camera with $2048 \times 2048$ pixels and a pixel scale of $5.5$
arcsec/pixel, resulting in $3.1 \times 3.1$ square degree field of
view (FOV). The readout time of the CCD is $9s$ and the saturation
level is reached with $16384$ ADU. In order to detect as many
photons as possible no filters are used. Guiding of
the telescope is performed by a $9$ cm aperture telescope equipped
with a ST-4 CCD camera.

The BEST system is located at the Observatoire de Haute-Provence
(OHP), France, since 2004. During the commissioning phase of the
BEST telescope for the OHP site, the system was upgraded to allow
remote control of all its components as well as a system and
environment monitoring. In autumn $2006$ the first observations in
remote control mode were performed by an observer at DLR in
Berlin. Since then, the system is controlled from Berlin with
occasional technical service visits at OHP. The remote observer
starts the system at the beginning of the night and operates until
the target field is acquired. Then, automatic exposure sequences are
taken throughout the night. The DLR staff cooperates with local
Berlin amateur astronomers who participate in the remote
observations.

Observations are obtained for typically 30,000 stars in a target field. The reduced stellar lightcurves are
archived for subsequent analysis.

\subsection{Data Center and Archiving}

In order to process the observational data acquired with the BEST telescope
sufficiently fast, a dedicated data processing center consisting of a computer
cluster system was installed at DLR during spring 2006. The data processing center
consists of 18 calculation nodes and 3 Tb disk space in total.

The BEST stellar lightcurves data base contains stellar coordinates, time
of observation, magnitude and magnitude error of the individual
measurements for each detected star within the BEST fields. To allow
a first estimate of stellar variability in the lightcurves a
variability index calculated according to \cite{Zhang 2003} has been
included.  The content of the data base is available to the scientific community upon request.

\section{Observations}

In the following we describe the data obtained on the CoRoT IRa01 and
LRc01 stellar fields which were used to check the transit candidates
of the later confirmed planets CoRoT-1b and CoRoT-2b.

\subsection{General Observing Modes}

The CCD is cooled to its operational temperature around $20^\circ$C below outdoor
temperature. Due to the location in Haute-Provence the outdoor temperature changed over 
the seasons with extreme values between $-10^\circ$C to $+20^\circ$C. However, variations 
during the observations were mostly limited to a couple of degrees once 
stabilized during the astronomical night. 
Calibration bias frames and dome flatfields are acquired at the beginning of each night. 

The general scientific observing sequence consists of a series of a
$40$ sec exposures followed by a $240$ sec exposure. Dark
frame images are taken every $25$ minutes with the same exposure
time as the science images. This fully automatized sequence takes
about $30$ minutes.  The resulting duty cycle for the scientific images on an average good night is around 50-60 \%.
The dark and bias frames acquired within the observing
session are used to monitor the performance of the CCD over the
night. At the end of the observing runs, additional bias and
dark frames are taken for calibration purposes.

\subsection{BEST Data Sets of the CoRoT Fields IRa01 and LRc01}

The IRa01 field was observed in $12$ nights spread over a period of
40 days from $11^{th}$ November to $21^{th}$ December
$2006$ \cite{Kabath07}. The LRc01 field was observed from $6^{th}$
June until $7^{th}$ Sept $2005$ in $35$ nights spread over $89$
days \cite{Karoff2007}. No observations were performed $6$ days around full Moon. 
Additional data gaps resulted from bad weather conditions at OHP.

The observed CoRoT IRa01 field is located at $RA=06^h 57^m 18^s$ and
$Dec=-01^\circ 42^{'} 00^{``}$ $(J2000.0)$ \cite{Michel06}. The BEST FOV covered
the CoRoT exoplanetary field only to avoid the bright stars in
the CoRoT asteroseismology field. The BEST FOV, therefore, was centered at
$RA=06^h 46^m 24^s$ and $Dec=-01^\circ 54^{'} 00^{``}$ $(J2000.0)$. 

The CoRoT LRc01 field is located at $RA=19^h
23^m 33^s$ and $Dec=+00^\circ 27^{'} 36^{``}$ \cite{Michel06}. Again, the BEST FOV
was offset to avoid the bright asteroseismology target stars and was centered at
$RA=19^h 00^m 00^s$ and $Dec=+00^\circ 01^{'} 55^{``}$.

Since the CoRoT fields are located near the celestial equator they can be
observed only at relatively low altitudes ($< 50$ degree) from OHP.
Therefore, the BEST data set had to be obtained at relatively high
airmass.  Typically observation arcs ranged between 3-6 hours per night.

In this paper we analyze the images acquired with $240$ seconds
exposure time since they provide an overlap with the CoRoT magnitude range with sufficient photometric precision.
The data sets contain in total $300$ frames for the IRa01
field and $727$ frames for the LRc01 field. The magnitude range of the
detected stars in the BEST fields  (about $30000$) is
$10$ mag to $18$ mag and thus overlaps with the CoRoT magnitude
range ($12-16$ mag). Typically, during a good photometric night BEST aquires lightcurves with a precision $\leq 1\%$ for 3000-4000 stars in the range $10-13$ mag.

\section{BEST Data Reduction Pipeline}

The BEST data reduction pipeline includes the basic CCD imaging reduction
steps, such as dark current and bias subtraction as well as flatfielding,
which have been performed by computing the related master frames for
each night (\cite{Rauer2004}, \cite{Voss06}). For the photometric data
reduction we used the ISIS image subtraction method.  A detailed description of the ISIS method can be found in \cite{Alard00}. 

In a first step, the stellar coordinates in every image are interpolated with respect to a single chosen reference frame from the middle of the observation run. For this purpose the matching routine interp from the ISIS package \cite{Alard00} was used. 
In this way slight translations are well corrected and a coordinate template common to all images is created.

In a second step, the ISIS image substraction method is applied to remove background contaminat 
to the stellar light. This is done by substracting a reference frame from all the
scientific frames thereby leaving only the remaining variability of each star as the final product.
In this process all frames have to match to the same seeing. 
The used reference frame is created from typically the $10$ best seeing images obtained during 
the whole observation campaign. This reference frame is thereafter subtracted from all other images, such that the 
stellar PSF on the reference frame is convolved with the corresponding stellar PSFs on the scientific images to best match the stellar PSFs in each individual frame. 
To obtain the flux variation of each star as a function of time,  aperture photometry 
is applied on the reference frame and 
on all convolved and substracted frames. The magnitude of each star was determined by aperture photometry in the reference frame. The magnitude at each time step was then determined by adding the deviations from the reference which were determined using the same aperture in the individual difference frames.
The implementation of the BEST data pipe line is described in \cite{Karoff2007}. 
Discussions  on the obtained photometric quality for the CoRoT fields 
IRa01 and LRc01 can be found in \cite{Kabath07} and \cite{Karoff2007}. A local extinction correction is performed within the ISIS package  
that significantly reduces the influence of differential 
extinction present due to the large FOV. We use a fixed aperture of $7$ pixels to match the FWHM 
range between $1-3$ pixels of the analyzed stars as discussed in \cite{Karoff2007}.

In the last step the (x,y) coordinates of the CCD frames are
transformed into right ascension and declination. For such
astrometric transformation we compared the $300$ brightest stars
in the BEST data set with the USNO-A2.0 catalog using the MATCH routine 
\cite{Valdes95} to determine the transformation parameters. 
Thereafter the transformation of the full BEST data set was performed and the 
corresponding sub-set of stars in the USNO-A2.0 identified.  A more detailed 
description of the approach can be found in \cite{Kabath07} 
and \cite{Karoff2007}. Using these stars we
calculated the shift in  magnitude with respect to the USNO-A2.0
catalog. Since only the relative magnitudes are needed to identify
the variable stars we do not intend to perform real absolute photometric
calibration.

The data can be affected by systematic errors due to zero point
offsets from night to night or large systematic trends due to
variable weather conditions. In order to remove such trends, we
apply a routine by \cite{Tamuz05} four times on both data sets. A more detailed analysis of the effect of systematic errors for the whole BEST data set is under investigation.

\section{Results}

\subsection{Photometric pre-discovery observations of CoRoT-1b and CoRoT-2b}

The two planets CoRoT-1b and CoRoT-2b were discovered by the CoRoT space mission in the "alarm mode" during observations of its "inital run" (IRa01) and the first "long run" (LRc01) fields (\cite{barge2008} and \cite{alonso2008}). The planets orbit relatively bright stars (V=13.6 mag for CoRoT-1b and V=12.57 mag for CoRoT-2b) but within the magnitude range covered by BEST. Furthermore, their transits are strong signals of more than 2 \% and 3\% depths, respectively. We therefore searched for signatures of the transit candidates in our BEST data archive, based on the determined transit epochs from the CoRoT alarm-mode operations. The star and planet parameters for CoRoT-1b and CoRoT-2b are given in 
Table \ref{tab:parameters}.

\begin{figure*}
\includegraphics[scale=1,angle=0,height=2cm]{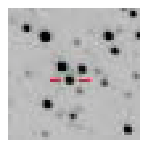} 
\includegraphics[scale=1,angle=0,height=2cm]{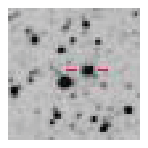}
\caption{Cut-outs (50$\times$50 pixels each\textbf{ (4.6$\times$4.6 arcmin)}) of the surroundings of CoRoT-1b (left) and CoRoT-2b (right) in BEST data. The position of the targets is marked. \label{stamps}}
\end{figure*}

The BEST archive data sets of the respective fields include 12 nights for the IRa01 and 35 nights for the LRc01 field. Figure \ref{stamps} shows the target stars and their near vicinity.
CoRoT-1b is located in a very crowded part of the CoRoT field. As a consequence, using the normal data pipeline with a photometric aperture of seven  pixels, resulted in a diluted signal where a neighbouring star contributed to the total flux within the aperture. To minimize this effect, manual photometry with an aperture size of five pixels was performed in order to obtain optimum photometric accuracy. In all other aspects the reduction for CoRoT-1b was performed as outlined in the previous section. CoRoT-2b was processed using the standard BEST pipeline.

\begin{figure*}
\includegraphics[scale=0.5,angle=0]{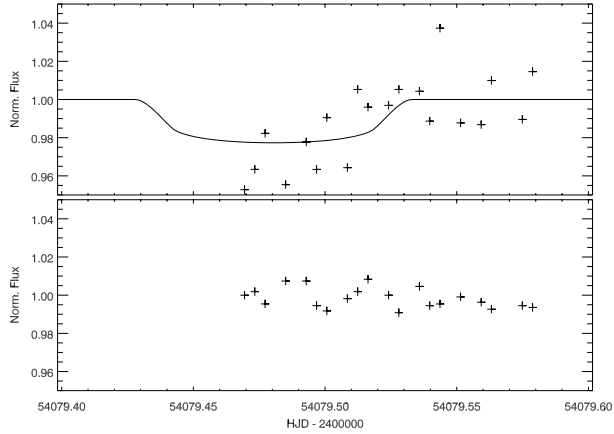}
\caption{Top panel: BEST archive lightcurve of a transit event of CoRoT-1b obtained on 10 December 2006. The solid line indicates the transit shape according to CoRoT data \cite{barge2008}. Lower panel: lightcurve of a nearby comparison star of similar magnitude. \label{exo1b}}
\end{figure*}

Due to the restricted data set, only one partial transit event of CoRoT-1b was covered by BEST observations. Figure \ref{exo1b} shows the lightcurve obtained for CoRoT-1b on 10 December 2006. Only the egress of the transit event could be covered at the beginning of the night. The data set is severely affected by noise due to the weather situation. In addition, the star is located  at the fainter end of the BEST magnitude range for signals of the CoRoT-1b order. Nevertheless the transit signature can be seen in the data. For comparison another star of similar magnitude in the vicinity is also presented in the figure. 

\begin{figure*}
\includegraphics[scale=0.5,angle=0]{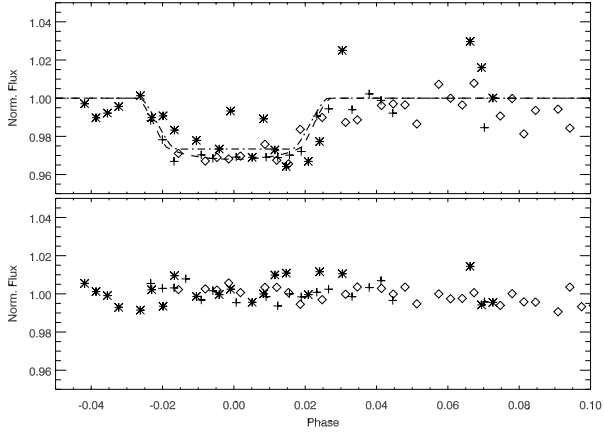}
\caption{Top panel: BEST archive composite lightcurve of three transit events of CoRoT CoRoT-2b obtained on 14 July 2005 ($+$), 28 July 2005 ($\star$) and 04 Aug 2005 ($\diamond$), respectively. Phase folding has been performed with $T_{c} [HJD] = 2454237.53562 $ and $P [d] = 1.7429964 $ determined from the CoRoT data \cite{alonso2008}. The "long dash" and "long-short dash" lines indicate the transit shape according to CoRoT data \cite{alonso2008} and an independent fitting of the BEST data respectively. The residuals of the lightcurve fitting (0.009 mag) are in good agreement with the RMS (0.01 mag) of the BEST lifgtcurves. Lower panel: lightcurve of a nearby comparison star. Both stars have a similar magnitude as observed with BEST. \label{exo2b}}
\end{figure*}

Figure \ref{exo2b} shows the composite lightcurve obtained on CoRoT-2b by BEST. In total, three transit events could be covered in the observing period from June to September 2005. Again, the data quality varies from night to night due to the weather situation, and on two nights only partial transits could be obtained. However, an almost complete transit signal is detected on 14 July 2005. Phase folding of the lightcurves from the individual nights has been done using the epochs and orbital period determined from the CoRoT data (see Table \ref{tab:parameters}). For comparison another star is also presented in the figure. A good agreement was found in orbital period and transit shape between the events detected by BEST and the more detailed CoRoT observations.

In both cases the pre-discovery observations of CoRoT-1b and CoRoT-2b with the BEST survey were used during CoRoT follow-up observations to refine early ephemeris and optimize the effort to confirm their planetary nature (\cite{barge2008} and \cite{alonso2008}). The lightcurve data used in this paper are given in Tables \ref{tab:Exo1bphot} and \ref{tab:Exo2bphot}.

\subsection{Detailed Analysis of a CoRoT-1b transiting event}

In our data set we found a partial transit event of CoRoT-1b (Figure \ref{exo1b}). The data points cover the egress event. This observation of a transit was obtained 60 days before the discovery observations by the CoRoT satellite. We therefore tried to constrain the midcenter time  to obtain an additional point for an O-C analysis.

For this purpose we fixed the stellar, planetary and orbital parameters at values given in \cite{barge2008}. The only free parameter was the midtime of the transit. The cycle number at the time of observations was $-53$, and therefore the predicted time of the transit is HJD 2~454~079.4785. \textbf{We scanned the region around this predicted time ($\pm$4 hours) with a step size of 0.5 seconds to find the best agreement between the model and the observed lightcurves.} Instead of the original light curve we used a smoothed light curve, applying a 2-points moving average\textbf{ in order to increase the SNR}. The resulting midcenter time of the transit was found to be $T_\mathrm{mid}\mathrm{(HJD)} = 2~454~079.4673 \pm 0.0088$.  The $O-C$ value corresponds to $O-C= -0.0112\pm0.0088$ days. The $O-C$ value determined from this partial light curve therefore is zero within the $3\sigma$ error bars, and we conclude that our observation is in agreement with the ephemeris of \cite{barge2008}. 

\subsection{Detailed Analysis of CoRoT-2b transiting events}

Our transit observations of CoRoT-2b were obtained $2$ years prior the
discovery, therefore we can extend its $O-C$ diagram which is the base of the
period studies.

The determination of the midtransit times can be done by fitting the light
curves of individual transit events, but there can be at least three
approaches: (1) fitting every parameters, (2) fitting every parameters except
the period, or (3) fitting only the midtransit times.
The first method - fitting all the parameters - could not be applied here
because we did not have enough data points to determine the epoch and period
only from our observations. The second and third method (fitting everything
except period which is fixed) were tried.

We chose the quadratic limb-darkened model of \cite{Mandel02} who gave 
analytic expressions for the light loss of the system. In this model the planet 
is dark without observable radiation while the stellar disc is limb-darkened 
and both objects are spherical. This model has seven free parameters: orbital 
inclination, radius of the planet and the size of the semi-major axis (both of 
them is expressed in stellar radius unit), two limb darkening coefficients, the 
period and the epoch. 

To find the best agreement between the model and the data a Markov Chain Monte 
Carlo method complemented with the Metropolitan-Hastings algorithm (see e.g. 
\cite{Croll06}) was used.

In Figure \ref{exo2b} we show the results of both fitting approaches. The "long dash" line
represents the model for the CoRoT data from Alonso et al. (2008), only the
midtransit time was adjusted. The "long-short dash line" represents the fitted
model to the BEST data where only the period was fixed at the value
given in Alonso et al. (2008).
In the descending and ascending branch of the  transit the agreement is perfect
but at the bottom part of the transit there is  a small deviation. This
deviation is caused by the different values of  limb-darkening coefficients.
These coefficients were fitted in \cite{alonso2008}  and in our case. The CoRoT
and BEST light curves have a different  number of data points and different
accuracies and hence we cannot determine  these coefficients from the BEST
lightcurves with such high accuracy as \cite{alonso2008}.  However, the
semi-major axis and planet radius are in reasonable good  agreement and the
inclination differs by only 2.5 degrees. This good agreement in  the
planet to star radii ratio and semi-major axis is due to the fact that the 
lightcurves at this photometric accuracy level are not too sensitive to the 
limb-darkening \cite{Southworth08}. This comparison  illustrates that
the data from  ground-based surveys can be used for a realistic estimation of
the planet parameters  of detected exoplanets.

\subsection{$O-C$ Analysis of CoRoT-2b}

Our observations of CoRoT-2b were obtained almost 2 years prior to the CoRoT
data.  We, therefore, investigated whether a significant O-C deviation could be
detected in comparison to the ephemerides given in  \cite{alonso2008}.
With these three new  transits in total  15 observed transit timing data
points are available for CoRoT-2b. The sources of the other 12 points are:  one
from \cite{alonso2008}, one from \cite{Veres09}, and ten from the AXA working
group (AXA:  Amateur eXoplanets Archive, see: brucegary.net/AXA/x.htm).

\begin{figure*}
\includegraphics[scale=1,angle=0]{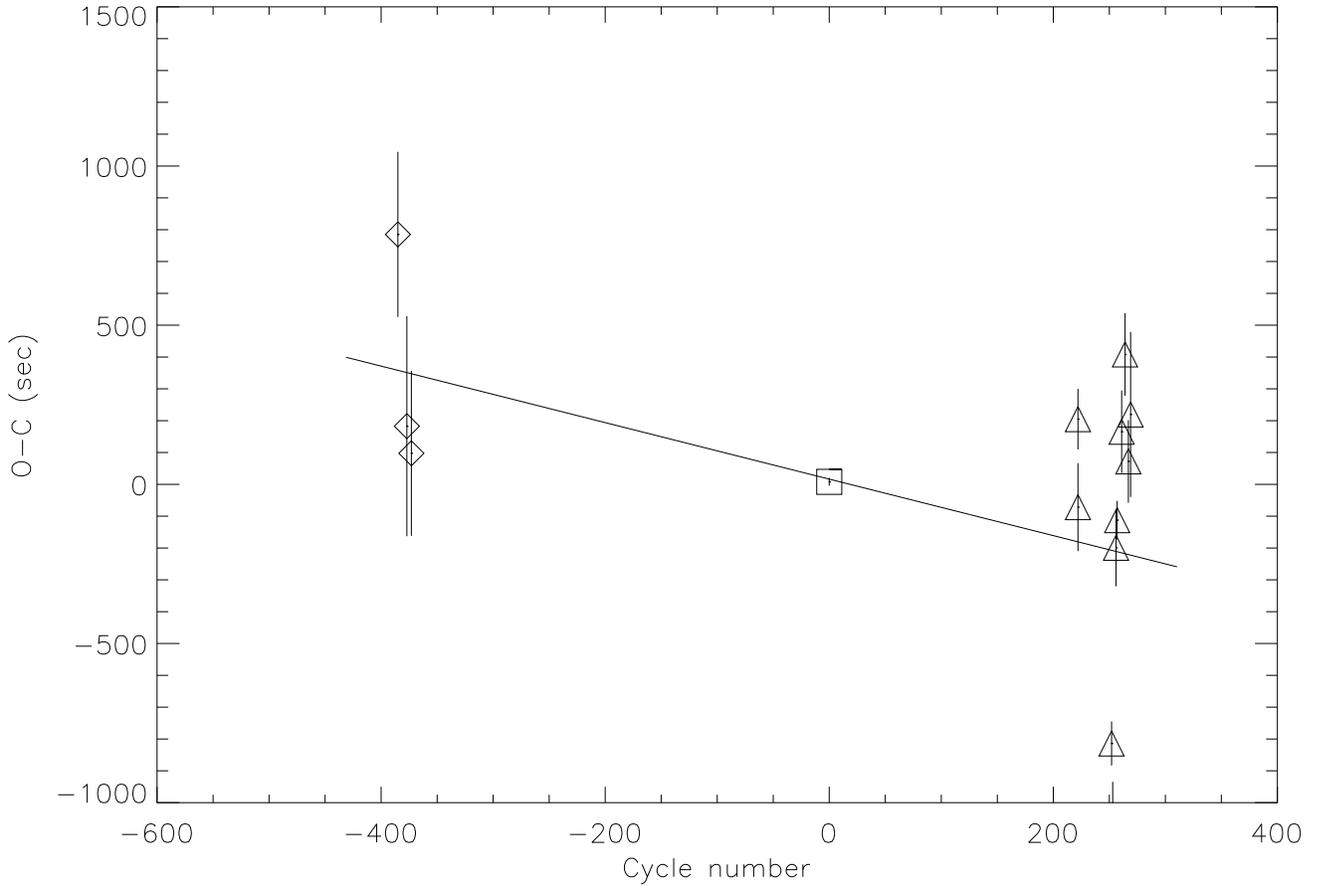}
\caption{The $O-C$ diagram of CoRoT-2b with $O-C$ vs. cycle number. Vertical lines indicate the error bars. ($\diamond$) represents the BEST meassurements, ($\Box$) the data CoRoT data from \cite{alonso2008}, while ($\triangle$) correspond to the observations from AXA and \cite{Veres09}. The solid line is a linear fit corresponding to Eq. 1. \label{oc_exo2b}}
\end{figure*}

 To determine the mid-time of the observed transits, we
fixed the model parameters  at the values given in \cite{alonso2008} with the
exception of the mid-time of the CoRoT transit. Only this parameter was
adjusted to obtain an optimized fit of all observed transiting events. The
corresponding $O-C$ values can be found in Table \ref{tab:epoch} and in
graphical form in Figure \ref{oc_exo2b}.

The interpretation of this diagram is difficult because of two reasons. First,
there is a lack of available  observational data, which means that the $O-C$ is
not well covered. Second, the precision of the published  minima times allows
only the study of large period changes.

The deviation of all $O-C$ values yielded $\chi^2 = 440$ when using the ephemeris of
Alonso et al. (2008). A simple linear fit of the $O-C$ diagram yielded the
following ephemeris:

\begin{equation}
T_{\mathrm{min}}~(\mathrm{HJD}) = 2454237.5357(16) + 1.7429861(58) \times E 
\end{equation}
With this ephemeris we reached $\chi^2 = 315$, and some improvement
in the $O-C$ deviations. The new period value (1.7429861 days) is shorter than
the one in \cite{alonso2008} by $0.9\pm0.5$ seconds.  We conclude that there
is no observable sign of period change in the available CoRoT-2b transit timing
data. The lack  of observed period change is in agreement with the
results of \cite{alonso2009}. They analysed the transit times in the CoRoT
data and found no clear sign of periodic period change which would have 
amplitudes larger than 10 seconds. However, \cite{alonso2009} did not publish their
individual $O-C$ data thereby preventing  a direct comparison with our result.
\cite{Veres09} did also not find any sign of period variation on a  shorter
baseline of observations.

\section{Summary}

We present pre-discovery observations of the first two planets detected by the CoRoT space mission, CoRoT-1b and CoRoT-2b. 
The transit events were detected in the BEST data archive, based on the epochs determined from the CoRoT data 
(\cite{barge2008}, \cite{alonso2008}). Although the observational duty cycle was severely affected by bad
weather conditions, partial transits of both planets were detected by BEST. A transit of CoRoT-1b was 
observed on 10 December 2006. Transits of CoRoT-2b were observed in 14 July 2005, 28 July 2005 and 04 Aug 2005. 
No significant O-C deviation in comparison to the ephemerides of \cite{alonso2008} was found.

When planetary candidates are first announced from the CoRoT alarm-mode, their ephemeris can be based on 
few transit events only. At this point, confirmation from ground-based observations taken one or more years 
prior to the spacecraft observations aids in checking the ephemeris quickly. In addition, these observations 
confirm the transit on the prime target and help identifying close possibly contaminating variable stars. 
We will continue to use the BEST data archive for this purpose in future to support the planet search in 
particular during the alarm-mode of CoRoT, when primarily deep transits around bright stars are detected 
which overlap with the detection range of BEST.

\begin{acknowledgements}  
We are grateful to the enthusiastic support of our
observations by Martin Dentel, Christopher Goldmann, Susanne Hoffman and Tino Wiese and
during the data reduction and archiving phase by Christopher Carl,
Sabrina Kirste and Mathias Wendt.
\end{acknowledgements}

\clearpage

\clearpage

\clearpage

\begin{center}
\begin{table}
\caption{The star and planet parameters for CoRoT-1b and CoRoT-2b. The limb darkening coefficients $u_{+}$ and $u_{-}$ are
defined according to \cite{brown01}. All parameters in the lower part of the table are from \cite{barge2008} and
\cite{alonso2008}.\label{tab:parameters}}
\begin{tabular}{c|c|cc}
			\tableline
			Parameters	& CoRoT-1b	&\multicolumn{2}{c}{CoRoT-2b}\\
			 &	\cite{barge2008}	& \cite{alonso2008} & This paper \\
			\tableline
			
			$a/R_{\star}$ & $4.92 \pm 0.08$ & $6.70 \pm 0.03$  & $6.32 \pm 0.48$  \\ 
			$R_{P}/R_{\star}$ & $0.1388 \pm 0.0021$ & $0.1667 \pm 0.0006$ & $0.1633 \pm 0.0031$  \\
			$i [deg]$ & $85.1 \pm 0.5$ & $87.84 \pm 0.10$  & $85.23 \pm 1.37$  \\
			$u_{+}$ & $0.71 \pm 0.16$ & $0.471 \pm 0.019$ & $-0.014 \pm 0.28$ \\
			$u_{-}$ & $0.13 \pm 0.30$ & $0.34 \pm 0.04$ & $0.024 \pm 0.2$ \\
			\tableline
			$RA [J2000.0]$ & $ 06^h 48^m 19.17^s$ & $ 19^h 27^m 06.5^s$ &  \\
			$Dec [J2000.0]$ & $-03^\circ 06^{'} 07.78^{``}$ & $01^\circ 23^{'} 01.5^{``}$ &  \\
			$V_{mag}$ & $13.6$ & $12.57$ &  \\			

			$P [d]$ & $1.5089557 \pm 0.0000064$ & $1.7429964 \pm 0.0000017$ &  \\
			$T_{c} [HJD]$ & $2454159.4532 \pm 0.0001$ & $2454237.53562 \pm 0.00014$ & \\
			\tableline
\end{tabular}
\end{table}
\end{center}

\begin{table*}
	\centering
		\begin{tabular}{c|c|c|c}
			\hline
			Transit midtime (HJD)	& N	& O-C (days) & Reference\\
			\hline			
				$54706.4041  \pm 0.0030$ &	$269$ &	$0.0024$ &	\cite{Veres09} \\
				$54702.9164  \pm 0.0015$ &	$267$ &	$0.0007$ &	AXA \\
				$54697.6913  \pm 0.0015$ &	$264$ &	$0.0046$ &	AXA \\
				$54692.4595  \pm 0.0015$ &	$261$ &	$0.0018$ &	AXA \\
				$54685.4843  \pm 0.0007$ &	$257$ &	$-0.0014$ &	AXA \\
				$54683.7403  \pm 0.0014$ &	$256$ &	$-0.0024$ &	AXA \\
				$54678.5016  \pm 0.0012$ &	$253$ &	$-0.0121$ &	AXA \\
				$54676.7612  \pm 0.0008$ &	$252$ &	$-0.0095$ &	AXA \\
				$54676.7577  \pm 0.0010$ &	$252$ &	$-0.0130$ &	AXA \\
				$54624.4831  \pm 0.0011$ &	$222$ &	$0.0023$ &	AXA \\
				$54624.4799  \pm 0.0016$ &	$222$ &	$-0.0009$ & AXA \\
				$54237.53562 \pm 0.00014$ &	$0$ &	$0.0000$ & \cite{alonso2008} \\
				$53587.399   \pm 0.003$ & $-373$ & $0.001$ & This paper \\
				$53580.430   \pm 0.004$ &   $-377$ & $0.004$ & This paper \\
				$53566.498   \pm 0.003$ &   $-385$ & $0.016$ & This paper \\
			\hline
		\end{tabular}
	\caption{Transit observations and $O-C$ deviation of CoRoT 2b. The
	cycle number (N) is relative to $T_{c} [HJD] = 2454237.53562 $
	determined from the CoRoT data \cite{alonso2008}.}
	\label{tab:epoch}
\end{table*}

\begin{center}
\begin{table}
\caption{Observed normalized intensities of CoRoT-1. Intensity is from white light magnitudes, normalized to
the out-of-eclipse light.\label{tab:Exo1bphot}}
\begin{tabular}{cc}
                     \tableline
HJD-2454000      & Intensity      \\
                        \tableline
			79.47007 &  0.957 \\
			79.47448 &  0.967 \\
			79.47886 &  0.986 \\
			79.48326 &  0.959 \\
			79.49129 &  0.982 \\
			79.49562 &  0.967 \\
			79.50198 &  0.994 \\
			79.50636 &  0.968 \\
			79.51075 &  1.009 \\
			79.51516 &  1.000 \\
			79.52320 &  1.001 \\
			79.52760 &  1.009 \\
			79.53714 &  1.008 \\
			79.54152 &  0.993 \\
			79.54592 &  1.041 \\
			79.55030 &  0.992 \\
			79.55828 &  0.991 \\
			79.56269 &  1.014 \\
			79.57540 &  0.994 \\
		 	79.57973 &  1.019 \\
                        \tableline
\end{tabular}
\end{table}
\end{center}

\small

\begin{center}
\begin{table}
\caption{ Observed normalized intensities of CoRoT-2. Intensity is from white light magnitudes, normalized to
the out-of-eclipse light.\label{tab:Exo2bphot}}
\begin{tabular}{cccc}
\tableline
HJD-2453000 & Intensity & HD-2453000 & Intensity    \\
                        \tableline
566.44083  & 0.995 & 580.46740  & 0.985     \\
566.44657  & 0.985 & 580.47851  & 1.032     \\
566.45225  & 0.974 & 580.54095  & 1.037\\
566.45798  & 0.760 & 580.54656  & 1.023      \\       
566.46555  & 0.977 & 580.55205  & 1.007    \\
566.47123  & 0.975 & 587.37038  & 0.979  \\
566.48264  & 0.976 &  587.38350  & 0.974\\
566.49729  & 0.976 & 587.38918  & 0.976  \\
566.50297  & 0.976 & 587.39491  & 0.976  \\
566.50870  & 0.977 & 587.40071  & 0.977  \\
566.51444  & 0.979 & 587.41267  & 0.983  \\
566.52201  & 0.998 & 587.41841  & 0.975   \\
566.52781  & 1.002 & 587.42415  & 0.973   \\
566.53928  & 1.001 & 587.42988  & 0.991   \\
566.54764  & 1.009 & 587.44063  & 0.997  \\
566.55344  & 1.006 &  587.45204  & 0.995 \\
566.55918  & 0.999 & 587.45784  & 0.996 \\
566.60392  & 0.992 &  587.46962  & 1.004   \\
580.35253  & 1.004 &  587.47535  & 1.005\\
580.35809  & 0.997 &  587.48109  & 1.004 \\
580.36364  & 0.999 &  587.48689  & 0.994 \\
580.36920  & 1.003 &   587.49763  & 1.015\\
580.37976  & 1.009 &   587.50337  & 1.007\\
580.38531  & 0.997 &   587.50911  & 1.004 \\
580.39086  & 0.998 &   587.51484  & 1.015\\
580.39642  & 0.991 &  587.52766  & 0.998\\
580.40716  & 0.985 &  587.53346  & 1.007\\
580.41833  & 0.980 & 587.53920  & 0.989\\
580.42388  & 1.001 &  587.54493  & 1.001\\
580.43444  & 0.976 &  587.55592  & 1.002 \\
580.43994  & 0.996 &  587.56166  & 0.991\\
580.44555  & 0.980 &   587.56739  & 0.828	\\
580.45110  & 0.971 & 587.57313  & 1.006 \\
580.46185  & 0.974 & & \\
\tableline
\end{tabular}
\end{table}
\end{center}

\end{document}